\documentclass[a4paper,11pt]{article} 
  
\usepackage[english]{babel}
\usepackage[a4paper]{geometry}
\usepackage{amsmath}
\usepackage{amsthm}
\usepackage{amssymb}
\usepackage{color}

\usepackage{hyperref}         

\def\bR{\mathbb{R}}
\def\bN{\mathbb{N}}
\def\bZ{\mathbb{Z}}

\def\cC{\mathcal{C}}

\def\cQ{\mathcal{Q}}
\def\cD{\mathcal{D}}

\def\cM{\mathcal{M}}
\def\cV{\mathcal{V}}
\def\cO{\mathcal{O}}
\def\cF{\mathcal{F}}
\def\cG{\mathcal{G}}
\def\cL{\mathcal{L}}
\def\cJ{\mathcal{J}}
\def\cN{\mathcal{N}}
\def\cE{\mathcal{E}}
\def\cK{\mathcal{K}}
\def\cH{\mathcal{H}}

\def\ph{\varphi}

\def\wt{\widetilde}

\let\g=\gamma     
\let\s=\sigma
\let\k=\kappa

\newtheorem{theorem}{Theorem}[section]  

\newtheorem{prop}[theorem]{Proposition}
\newtheorem{lemma}[theorem]{Lemma}

\numberwithin{equation}{section}

\begin{document}
\title{Bogoliubov Excitation Spectrum for \\ Bose-Einstein Condensates}

\author{Benjamin Schlein \\
\\
Institute of Mathematics, University of Zurich\\
Winterthurerstrasse 190, 8057 Zurich, 
Switzerland}

\maketitle

\begin{abstract}
We consider interacting Bose gases trapped in a box $\Lambda = [0;1]^3$ in the Gross-Pitaevskii limit. Assuming the potential to be weak enough, we establish the validity of Bogoliubov's prediction for the ground state energy and the low-energy excitation spectrum. These notes are based on \cite{BBCS3}, a joint work with C. Boccato, C. Brennecke and S. Cenatiempo.
\end{abstract}

\section{Introduction}

In the last two decades, since the first experimental realisations of Bose-Einstein condensates \cite{BEC1,BEC2}, the study of bosonic systems at low temperature has been a very active field of research in physics (experimental and theoretical) and also in mathematics. 

Trapped Bose gases observed in typical experiments are well described as quantum systems of $N$ particles, interacting through a repulsive two-body potential with scattering length of the order $N^{-1}$; this asymptotic regime is commonly known as the Gross-Pitaevskii limit. If particles are confined in a  
box $\Lambda = [0;1]^3$ and we impose periodic boundary conditions, the Bose gas in the Gross-Pitaevskii regime is described by the Hamilton operator 
\begin{equation}\label{eq:ham0} H_N = \sum_{j=1}^N -\Delta_{x_j} + \kappa \sum_{i<j}^N N^2 V(N (x_i - x_j)) \, . \end{equation}
According to the bosonic statistics, (\ref{eq:ham0}) acts on the Hilbert space $L^2 (\Lambda)^{\otimes_s N}$, the subspace of $L^2 (\Lambda^N)$ consisting of all functions that are invariant with respect to permutations of the $N$ particles. In (\ref{eq:ham0}), $V$ is assumed to be non-negative, spherically symmetric, compactly supported and sufficiently regular (in fact, the condition $V \in L^3 (\bR^3)$ will suffice) and $\kappa > 0$ is a coupling constant that will be later supposed to be small enough (but fixed, independent of $N$).  We denote by $\frak{a}_0$ the scattering length of $\kappa V$, which is defined by the requirement that the solution of the zero-energy scattering equation  
\begin{equation}\label{eq:scat} \left[-\Delta + \frac{\kappa}{2}V (x) \right] f (x) = 0 \end{equation}
with the boundary condition $f (x) \to 1$ as $|x| \to \infty$, is given by $f(x) = 1 - \frak{a}_0 / |x|$, for $|x|$ large enough (outside the range of $V$). Equivalently, we can determine the scattering length through 
\begin{equation}\label{eq:scat2} 8 \pi \frak{a}_0 = \kappa \int V (x) f (x) dx \, . \end{equation}

It follows from the results of Lieb-Yngvason \cite{LY} and of Lieb-Seiringer-Yngvason \cite{LSY} that 
the ground state energy $E_N$ of (\ref{eq:ham0}) is such that \begin{equation}\label{eq:gs0} 
\lim_{N \to \infty} \frac{E_N}{N} = 4\pi \frak{a}_0 \, .  \end{equation}
Furthermore, the work of Lieb-Seiringer \cite{LS}, recently revised also in \cite{NRS}, implies that the ground state of (\ref{eq:ham0}) exhibits Bose-Einstein condensation. In other words, if $\psi_N \in L^2 (\Lambda)^{\otimes_s N}$ is the normalized ground state of (\ref{eq:ham0}) and $\gamma_N$ denotes the one-particle reduced density associated with $\psi_N$, which is defined as the non-negative trace class operator on $L^2 (\Lambda)$ with the integral kernel 
\[ \gamma_N (x;y) = \int dx_2 \dots dx_N \, \psi_N (x, x_2, \dots , x_N) \, \overline{\psi_N} (y, x_2, \dots , x_N) \]
then, as $N \to \infty$,  
\begin{equation}\label{eq:cond0} \gamma_N \to |\ph_0 \rangle \langle \ph_0| \end{equation}
where $\ph_0 (x) = 1$ for all $x \in \Lambda$ is the zero-momentum mode. The convergence (\ref{eq:cond0}) (which holds in any reasonable topology, for example with respect to the trace-class norm) means that, in the ground state of (\ref{eq:ham0}), all particles are described by $\ph_0$, up to a fraction vanishing in the limit of large $N$. One should however stress the fact that (\ref{eq:cond0}) does not imply that the product state $\ph_0^{\otimes N}$ is a good approximation for the ground state of (\ref{eq:ham0}). In fact, a simple computation shows that 
\begin{equation}\label{eq:fact} \langle \ph_0^{\otimes N} , H_N \ph_0^{\otimes N} \rangle = \frac{(N-1)}{2}  \kappa \widehat{V} (0) \end{equation}
which is not compatible with (\ref{eq:gs0}). The point, which will often recur in these notes, is that, because of the singular interaction, the ground state of (\ref{eq:ham0}) (and, in fact, all low-energy states, as we will see below) develops a short scale correlation structure, varying on the length scale $N^{-1}$ (and therefore disappearing in the limit $N \to \infty$), which is responsible for lowering the energy from (\ref{eq:fact}) to (\ref{eq:gs0}) (from (\ref{eq:scat2}) it is clear that $8\pi \frak{a}_0 < \kappa \widehat{V} (0)$).

Eq. (\ref{eq:gs0}) establishes $E_N$ to leading order. Our goal in these notes is to obtain more precise information about the ground state energy and about low-energy excitations of (\ref{eq:ham0}), determining them up to errors that vanish in the limit $N \to \infty$. 
\begin{theorem}\label{thm:main}
Let $V \in L^3 (\bR^3)$ be non-negative, spherically symmetric and compactly supported. Let the coupling constant $\kappa > 0$ be small enough.  Then we have
\begin{equation}\label{eq:gs2}
\begin{split} 
E_{N} = \; &4\pi (N-1) \frak{a}_N \\ &-\frac{1}{2}\sum_{p\in \Lambda^*_+}  \left[ p^2+8\pi \frak{a}_0 - \sqrt{|p|^4 + 16 \pi \frak{a}_0 p^2} - \frac{(8\pi \frak{a}_0)^2}{2p^2}\right] + \cO (N^{-1/4})
    \end{split}
    \end{equation}
where $\Lambda^*_+ = 2\pi \bZ^3 \backslash \{ 0 \}$, and 
    \begin{equation} \label{bN}
    \begin{split}
    8\pi \frak{a}_N = \; & \kappa \widehat{V} (0)
\\ &+\sum_{k=1}^{\infty}\frac{(-1)^{k} \kappa^{k+1}}{(2N)^{k}} \sum_{p_1, \dots, p_{k}\in \Lambda^*_+} \frac{\widehat{V} (p_1 /N)}{p_1^2} \left(\prod_{i=1}^{k-1}\frac{\widehat{V} ((p_i-p_{i+1})/N)}{p^2_{i+1}}\right)  \widehat{V} (p_{k}/N) \, . 
    \end{split}
    \end{equation}
Furthermore, the spectrum of $H_N-E_{N}$ below a threshold $\zeta > 0$ consists of eigenvalues given, in the limit $N \to \infty$, by 
    \begin{equation}
    \begin{split}\label{eq:exc}
    \sum_{p\in\Lambda^*_+} n_p \sqrt{|p|^4+ 16 \pi \frak{a}_0 p^2}+ \cO (N^{-1/4} (1+ \zeta^3)) \, .
    \end{split}
    \end{equation}
Here $n_p \in \bN$ for all $p\in\Lambda^*_+$ and $n_p \not = 0$ for finitely many $p\in \Lambda^*_+$ only.  
\end{theorem}
{\it Remarks:} 
\begin{itemize}
\item[1)] Taylor expanding the square root to third order, it is easy to check that the sum 
in (\ref{eq:gs2}) is absolutely convergent and therefore that it gives a contribution of order 
one to the ground state energy $E_N$. 
\item[2)] The $k$-th term in (\ref{bN}) is bounded by $C^k \kappa^k$, for some constant $C >0$. Hence, the series is absolutely convergent and bounded uniformly in $N$, if $\kappa > 0$ is small enough. 
\item[3)] The expression (\ref{bN}) for $8\pi \frak{a}_N$ can be compared with the Born series for the unscaled scattering length $\frak{a}_0$, given by 
 \begin{equation}\label{eq:flBorn}
        \begin{split}
        8\pi \frak{a}_0=\; & \kappa \widehat{V} (0) + \sum_{k=2}^{\infty}\frac{(-1)^{k} \kappa^{k+1}}{2^{k}(2\pi)^{3k}}\int_{\mathbb{R}^{3k}}dp_1 \dots  dp_k \, \frac{\widehat{V} (p_1)}{p_1^2} \left(\prod_{i=1}^{k-1}\frac{\widehat{V} (p_i-p_{i+1})}{p^2_{i+1}}\right)  \widehat{V} (p_{k}).
        \end{split}
        \end{equation}
In particular, it is possible to show that the difference $4 \pi (\frak{a}_N - \frak{a}_0) N$ remains bounded, of order one, in the limit of large $N$. Notice, however, that it does not seem to tend to zero; this means that, in (\ref{eq:gs2}), we cannot replace $\frak{a}_N$ with $\frak{a}_0$. In other words, at the level of precision of (\ref{eq:gs2}), the ground state energy is sensitive to the finite size of the box and it cannot be simply expressed in terms of the infinite volume scattering length $\frak{a}_0$. 
\end{itemize}
The results of Theorem \ref{thm:main}, in particular the expression (\ref{eq:exc}) for the excitation spectrum, have already been predicted by Bogoliubov in \cite{B}. In his work, Bogoliubov rewrote the Hamilton operator (\ref{eq:ham0}) as
\begin{equation}\label{eq:HN-bog} H_N = \sum_{p \in \Lambda^*} p^2 a_p^* a_p + \frac{\kappa}{2N} \sum_{p,q,r \in \Lambda^*} \widehat{V} (r/N) a_{p+r}^* a_q^* a_p a_{q+r} \end{equation}
using the formalism of second quantization. For every momentum $p \in \Lambda^* = 2\pi \bZ^3$, $a_p^*, a_p$ are the usual creation and annihilation operators defined on the bosonic Fock space $\cF = \bigoplus_{n \geq 0} L^2 (\Lambda)^{\otimes_s n}$ and satisfying canonical commutation relations $[a_p , a_q^*] = \delta_{pq}$, $[a_p, a_q ] =[a^*_p, a^*_q] = 0$. Since low-energy states exhibit Bose-Einstein condensation, we expect the operator $a_0^* a_0$ measuring the number of particles in the zero-momentum state $\ph_0$ to be of the order $N$ and therefore much larger than the commutator   
$[a_0, a_0^* ] =1$. Motivated by this observation, Bogoliubov decided to replace, in (\ref{eq:HN-bog}), all creation and annihilation operators $a_0^*, a_0$ by $\sqrt{N}$ and then to neglect all resulting terms with more than two creation and annihilation operators associated with momenta different than zero. With this approximation, Bogoliubov derived an Hamilton operator quadratic in creation and annihilation operators $a_p^* , a_p$ with $p \not = 0$ that he could diagonalize explicitly. Finally he argued, following a hint of Landau, that certain expressions that appeared in his formulas for the ground state energy and for the excitation spectrum were just first and second order Born approximations of the scattering length, and thus he replaced them with $\frak{a}_0$; with this final substitution he obtained essentially results equivalent to those stated in Theorem \ref{thm:main}. 

From the point of view of mathematical physics, the validity of the Bogoliubov approximation has been first established by Lieb-Solovej \cite{LSo} in the computation of the ground state energy of the one-component charged Bose gas. It was then proved by Giuliani-Seiringer \cite{GiuS} in their derivation  of the Lee-Huang-Yang formula for the ground state energy of a Bose gas in a combined weak coupling and high density regime, and by Seiringer \cite{Sei}, Grech-Seiringer \cite{GS}, Lewin-Nam-Serfaty-Solovej \cite{LNSS}, Derezinski-Napiokowski \cite{DN}, Pizzo \cite{P3} in their analysis of the low-energy spectrum of Bose gases in the mean field limit. More recently, the validity of Bogoliubov prediction was established in \cite{BBCS2} for systems of $N$ bosons interacting through singular potential, described by the Hamiltonian (written like (\ref{eq:HN-bog}) in second quantized form) 
\begin{equation}\label{eq:HNbeta} H^\beta_N = \sum_{p \in \Lambda^*} p^2 a_p^* a_p + \frac{\kappa}{2N} \sum_{p,q,r \in \Lambda^*} \widehat{V} (r/N^\beta) a_{p+r}^* a_q^* a_p a_{q+r} \end{equation}
for a parameter $\beta \in (0;1)$. Notice that (\ref{eq:HNbeta}) interpolates between the mean-field regime that is recovered for $\beta = 0$ and the Gross-Pitaevskii limit, which corresponds to 
$\beta =1$. 

In the rest of these notes, we are going to sketch the main ideas going into the proof of Theorem \ref{thm:main}; for more details, see \cite{BBCS3}. 

\section{Excitation Hamiltonian} 

The first step in the proof of Theorem \ref{thm:main} consists in factoring out the condensate to focus on its orthogonal excitations. We use here an idea from \cite{LNSS}. Since we expect that, for low-energy states, most particles occupy the zero-momentum mode $\ph_0 (x) = 1$ for all $x \in \Lambda$ (for the ground state, this follows from (\ref{eq:cond0})), we write an arbitrary $N$-particle wave function $\psi \in L^2 (\Lambda)^{\otimes_s N}$ as   
\[ \psi_N = \alpha_0 \, \ph_0^{\otimes N} + \alpha_1 \otimes_s \ph_0^{\otimes (N-1)} +\alpha_2 \otimes_s \ph_0^{\otimes (N-2)} +  \dots + \alpha_N \]
where $\alpha_j \in L^2_{\perp} (\Lambda)^{\otimes_s j}$, for all $j=0,1, \dots , N$. Here, $L^2_\perp (\Lambda)$ denotes the orthogonal complement of the one-dimensional subspace spanned by $\ph_0$ in $L^2 (\Lambda)$. It is easy to check that the choice of $\alpha_0, \dots , \alpha_N$ is unique, and that $\sum_{j=0}^N \| \alpha_j \|^2 = \| \psi_N \|^2$. Hence, with the notation 
\[ \cF_+^{\leq N} = \bigoplus_{j=0}^N  L^2_\perp (\Lambda)^{\otimes_s j} \]
for the truncated Fock space constructed over $L^2_\perp (\Lambda)$, we can define 
a unitary map \[ \begin{array}{llll} U_N : &L^2 (\Lambda)^{\otimes_s N} &\to &\cF_+^{\leq N} \\ &\psi_N  &\to & \{ \alpha_0 , \dots , \alpha_N \} \, . \end{array}   \]

The map $U_N$ allows us to focus on the orthogonal excitations of the condensate that are described in the Hilbert space $\cF_+^{\leq N}$. Conjugating the Hamiltonian (\ref{eq:ham0}) with the unitary map $U_N$, we define an excitation Hamiltonian $\cL_N = U_N H_N U_N^* : \cF_+^{\leq N} \to \cF_+^{\leq N}$. With the notation $\cN_+$ for the number of particles operator on $\cF_+^{\leq N}$, we find 
\begin{equation}\label{eq:rules} \begin{split} U_N a_p^* a_q U^*_N &= a_p^* a_q \, , \\
U_N a_p^* a_0 U^*_N &= a_p^* \sqrt{N - \cN_+} \, , \\
U_N a_0^* a_p U^*_N &= \sqrt{N- \cN_+} a_p \, , \\
U_N a_0^* a_0 U^*_N &= (N-\cN_+ ) \, .  \end{split} \end{equation}
Applying these rules to the Hamiltonian (\ref{eq:ham0}) written in second quantized form as in  (\ref{eq:HN-bog}), we arrive at \begin{equation}\label{eq:cLN} \cL_N = \cL_N^{(0)} + \cL_N^{(2)} + \cL_N^{(3)} + \cL_N^{(4)},\end{equation} 
with (recall the notation $\Lambda^*_+ = 2\pi \bZ^3 \backslash \{ 0 \}$) 
\begin{equation}\label{eq:cLNj}  \begin{split} 
\cL_{N}^{(0)} =\;& \frac{N-1}{2N} \kappa\widehat{V} (0) (N-\cN_+ ) + \frac{\kappa\widehat{V} (0)}{2N} \cN_+  (N-\cN_+ ) \, ,  \\
\cL^{(2)}_{N} =\; &\sum_{p \in \Lambda^*_+} p^2 a_p^* a_p + \sum_{p \in \Lambda_+^*} \kappa\widehat{V} (p/N) a_p^* a_p \left(\frac{N-\cN_+}{N}\right)  \\ &+ \frac{\kappa}{2} \sum_{p \in \Lambda^*_+} \widehat{V} (p/N) \left[ a_p^* a_{-p}^* \sqrt{\frac{N-1-\cN_+}{N} \frac{N-\cN_+}{N}} + \text{h.c.} \right]  \, ,  \\
\cL^{(3)}_{N} =\; &\frac{\kappa}{\sqrt{N}} \sum_{p,q \in \Lambda_+^* : p+q \not = 0} \widehat{V} (p/N) \left[ a^*_{p+q} a^*_{-p} a_q \sqrt{\frac{N-\cN_+}{N}}  + \text{h.c.}  \right]  \,  , \\
\cL^{(4)}_{N} =\; & \frac{\kappa}{2N} \sum_{p,q \in \Lambda_+^*, r \in \Lambda^*: r \not = -p,-q} \widehat{V} (r/N) a^*_{p+r} a^*_q a_p a_{q+r} \, , 
\end{split} \end{equation}
where $\text{h.c.}$ indicates the hermitian conjugate operator and where, in the notation $\cL^{(j)}_N$, the label $j \in \{ 0 ,2,3,4 \}$ refers to the number of creation and annihilation operators.

Conjugation with $U_N$ extracts  contributions from the quartic interaction in (\ref{eq:HN-bog}) and moves them into the constant and the quadratic parts $\cL_N^{(0)}$ and $\cL^{(2)}_N$ of the excitation Hamiltonian.  In the mean-field case considered in \cite{Sei,GS,LNSS,DN,P3} (corresponding to the Hamiltonian (\ref{eq:HNbeta}) with $\beta = 0$), one can show that, after application of $U_N$, the cubic and quartic terms $\cL^{(3)}_N$ and $\cL^{(4)}_N$ are negligible on low-energy states, in the 
limit $N \to \infty$. In this case, the low-lying excitation spectrum can therefore be determined diagonalizing the quadratic operator $\cL^{(2)}_N$. This is not the case in the Gross-Pitaevskii regime considered here. Applying the unitary map $U_N$ we factor out the condensate but we do not remove the short scale correlation structure which, as explained after (\ref{eq:fact}), still carries an energy of order $N$. As a consequence, in the Gross-Pitaevskii regime, cubic and quartic terms in $\cL_N$ are not negligible on low-energy states. 

Notice that conjugation with $U_N$ can be interpreted as a rigorous version of the substitution proposed by Bogoliubov of all creation and annihilation operators $a_0^*, a_0$ associated with zero momentum with factors of $\sqrt{N}$. The fact that $\cL^{(3)}_N$ and $\cL^{(4)}_N$ are not negligible means, therefore, that in the Gross-Pitaevskii regime the Bogoliubov approximation cannot be justified. But then, why did Bogoliubov obtained the correct expressions for the low-energy spectrum, the same expressions appearing in Theorem~\ref{thm:main}? The point is that, when at the end of his computation Bogoliubov replaced, following the hint of Landau, first and second Born approximations with the full scattering length $\frak{a}_0$, he exactly made up for the (non-negligible) contributions that are hidden in $\cL^{(3)}_N$ and $\cL^{(4)}_N$ and that he neglected with his approximation. 

It is clear that to obtain a rigorous proof of Theorem \ref{thm:main} we cannot neglect cubic and quartic parts of the excitation Hamiltionian. Instead, to extract the important contributions from $\cL^{(3)}_N$ and $\cL^{(4)}$, we need to conjugate $\cL_N$ with another unitary map, a map that implements  correlations among particles. 

\section{Generalized Bogoliubov Transformations} 

A strategy to implement correlations has been introduced in \cite{BDS}, a paper devoted to the study of the dynamics in the Gross-Pitaevskii regime, for approximately coherent initial data in the bosonic 
Fock space. In that paper, correlations were produced by unitary conjugation with a Bogoliubov transformation of the form
\begin{equation}\label{eq:Bog1} \wt{T} (\eta) = \exp \left[ \frac{1}{2} \sum_{p \in \Lambda^*_+} \eta_p (a_p^* a_{-p}^* - a_p a_{-p}) \right] \end{equation}
for an appropriate real function $\eta \in \ell^2 (\Lambda^*_+)$ (in fact, in \cite{BDS} the problem is not translation invariant and therefore slightly more complicated transformations were considered). Bogoliubov transformations are very convenient because their action on creation and annihilation operators is explicitly given by
\begin{equation} \label{eq:expl} \wt{T}^* (\eta) \, a_q \, \wt{T} (\eta)  = \cosh (\eta_q) \, a_q + \sinh (\eta_q) \, a_{-q}^* \, . \end{equation}
Unfortunately, Bogoliubov transformations of the form (\ref{eq:Bog1}) do not preserve the number of particles and therefore they do not leave the excitation Hilbert space $\cF_+^{\leq N}$ invariant. To solve this problem, we follow \cite{BS} and we introduce, on $\cF_+^{\leq N}$, modified creation and annihilation operators defined, for any $p \in \Lambda_+^*$, by
\[ b_p^* = a_p^* \, \sqrt{\frac{N-\cN_+}{N}} , \qquad \text{and } \qquad b_p = \sqrt{\frac{N-\cN_+}{N}} \, a_p \, . \]
Observing that, from (\ref{eq:rules}), 
\begin{equation}\label{eq:UbU} U^*_N \, b_p^* \, U_N = a_p^* \, \frac{a_0}{\sqrt{N}} , \qquad U^*_N \, b_p \, U_N = \frac{a_0^*}{\sqrt{N}} \, a_p  \, , \end{equation}
we conclude that the modified creation operator $b_p^*$ creates a particle with momentum $p$ and, at the same time, it annihilates a particle from the condensate (i.e. a particle with momentum $p=0$) while $b_p$ annihilate a particle with momentum $p$ and creates a particle in the condensate. In other words, $b_p^*$ creates and $b_p$ annihilates an excitation with momentum $p$, preserving however the total number of particles. This is the reason why modified creation and annihilation operators leave the excitation Hilbert space $\cF_+^{\leq N}$ invariant, in contrast with the standard creation and annihilation operators.  

Using the modified field operators we can now introduce generalized Bogoliubov transformations by defining, in analogy to (\ref{eq:Bog1}),   
\begin{equation}\label{eq:gen-Bog}
T (\eta) =  \exp \left[ \frac{1}{2} \sum_{p \in \Lambda^*_+} \eta_p (b_p^* b_{-p}^* - b_p b_{-p}) \right] \, . \end{equation}
By construction, $T(\eta) : \cF_+^{\leq N} \to \cF_+^{\leq N}$. The price we have to pay for replacing the original Bogoliubov transformations (\ref{eq:Bog1}) with their generalization (\ref{eq:gen-Bog}) is the fact that there is no explicit formula like (\ref{eq:expl}) describing the action of $T(\eta)$ on creation and annihilation operators (because modified creation and annihilation operators do not satisfy canonical commutation relations). Still, when we consider states exhibiting Bose-Einstein condensation where $a_0, a_0^* \simeq \sqrt{N}$, we may expect from (\ref{eq:UbU}) that $b_p \simeq a_p$ and $b_p^* \simeq a_p^*$ and therefore that (\ref{eq:expl}) is approximately correct, even if we replace $\wt{T} (\eta)$ by $T(\eta)$. It is possible to quantify this last statement through the introduction of remainder operators. For $p \in \Lambda^*_+$, we define $d_p, d^*_p$ by 
\begin{equation}\label{eq:TbT} \begin{split} T^* (\eta) \, b_p \, T(\eta) &= \cosh (\eta_p) \, b_p + \sinh (\eta_p) \, b^*_{-p} + d_p , \\  
T(\eta) \, b^*_p \, T(\eta) &= \cosh (\eta_p) \, b^*_p + \sinh (\eta_p) \, b_{-p} + d^*_p \, . \end{split} \end{equation}
Then it is possible to prove that, if $\eta \in \ell^2 (\Lambda^*_+)$ with $\| \eta \|_2$ small enough, 
\begin{equation}\label{eq:dp} 
\begin{split} 
 \| d^*_p \, \xi \| &\leq \frac{C}{N} \| (\cN_+ + 1)^{3/2} \xi \|, \\  \| d_p \, \xi \| &\leq \frac{C}{N} \| (\cN_+ + 1)^{3/2} \xi \| \, , \end{split} \end{equation}
for all $\xi \in \cF^{\leq N}_+$. On states exhibiting condensation, the operator $\cN_+$ is small; in this case (\ref{eq:dp}) can be use to show that the remainder operators $d_p, d_p^*$ are small (we gain a factor $N^{-1}$). The bounds (\ref{eq:dp}) (and some more refined version) are discussed in \cite[Section 7]{BBCS3}; their proof is based on \cite[Lemma 2.5]{BBCS1} which is a translation to momentum space of \cite[Lemma 3.2]{BS}. 

To implement correlations, the choice of the coefficients $\eta_p$ in (\ref{eq:gen-Bog}) must be  
related with the solution of the zero-energy scattering equation (\ref{eq:scat}). More precisely, since 
we are working on the finite box $\Lambda = [0;1]^{3}$, we consider the Neumann problem 
\begin{equation}\label{eq:neumann} \left[ -\Delta + \frac{\kappa}{2} V (x) \right] f_\ell (x) = \lambda_\ell f_\ell (x) \end{equation}
on the ball $|x| \leq N \ell$ with the normalization $f_\ell (x) = 1$ on the boundary $|x| = N\ell$. We find that the smallest Neumann eigenvalue $\lambda_\ell$ is such that 
\[ \lambda_\ell = \frac{3\frak{a}_0}{N^3 \ell^3} \left[ 1 + \cO \left( \frac{\frak{a}_0}{N\ell} \right) \right] \]
and that $f_\ell = 1- w_\ell$, where   
\begin{equation}\label{eq:well-bds} 0 \leq w_\ell (x) \leq \frac{C\kappa}{|x|+1} , \qquad |\nabla w_\ell (x)| \leq \frac{C\kappa}{|x|^2 + 1} \end{equation}
for all $|x| \leq N\ell$ (this confirms the intuition that $f_\ell$ is a small modification of the solution of the zero energy scattering equation (\ref{eq:scat}) which is given, for large $|x|$, by $1- \frak{a}_0/|x|$). 
By scaling, we find that 
\[ \left[ -\Delta + \frac{\kappa N^2}{2} V (N.) \right] f_\ell (N.) = \lambda_\ell N^2 f_\ell (N.) \]
on the ball $|x| \leq \ell$. Fixing $\ell < 1/2$ (independently of $N$), we can extend $f_\ell (Nx) = 1$ and also $w_\ell (Nx) = 1- f_\ell (Nx) = 0$ for all $x \in \Lambda$ with $|x| > \ell$. Hence, the maps $x \to w_\ell (Nx)$ and $x \to f_\ell (Nx)$ can be expressed as Fourier series with coefficients $N^{-3} \widehat{w}_\ell (p/N)$ and, respectively, $\delta_{p,0} - N^{-3}  \widehat{w}_\ell (p/N)$, for all $p \in \Lambda^*$. Here 
\[ \widehat{w}_\ell (z) = \frac{1}{(2\pi)^3} \int dx \, e^{-i x\cdot z} w_\ell (x) \]
is the Fourier transform of $w_\ell$ (as a compactly supported function on $\bR^3$). For $p \in \Lambda^*$, we define
\begin{equation}\label{eq:etap} \eta_p = - \frac{1}{N^2} \widehat{w}_\ell (p/N) \, .  \end{equation}
From (\ref{eq:well-bds}), it is easy to check that \begin{equation}\label{eq:etap-l2} |\eta_p| \leq C \kappa |p|^{-2}\end{equation} for all $p \in \Lambda^*_+$. It follows that $\eta \in \ell^2 (\Lambda^*_+)$, with $\| \eta \|_2 \leq C$, uniformly in $N$. On the other hand, it is important to notice that (\ref{eq:etap-l2}) does not provide enough decay in momentum to estimate the $H^1$-norm of $\eta$. Since the decay of $\widehat{w}_\ell (p/N)$  kicks in for $|p| \gtrsim N$, we obtain that \begin{equation}\label{eq:eta-H1} \| \eta \|^2_{H^1}   = \sum_{p \in \Lambda^*_+} (1+p^2) |\eta_p|^2 \simeq C N \, . \end{equation}

From (\ref{eq:TbT}) and using the notation $\cK = \sum_{p \in \Lambda^*_+} p^2 a_p^* a_p$ for the kinetic energy operator, it is easy to check that 
\begin{equation}\label{eq:TNKT} \begin{split} T^* (\eta) \,  \cN_+ \, T(\eta) &\simeq \cN_+ + \| \eta \|_2^2 \, , \\  T^* (\eta) \, \cK \, T(\eta) &\simeq \cK + \| \eta \|_{H^1}^2 \, . \end{split} \end{equation}
The uniform bound for $\| \eta \|_2$ and the estimate (\ref{eq:eta-H1}) for the $H^1$-norm of $\eta$ imply, therefore, that conjugation with $T(\eta)$ only creates finitely many excitations of the condensates, but also that these excitations carry a macroscopic energy, of order $N$ (in (\ref{eq:TNKT}) we only consider the change of the kinetic energy but also the change of the potential energy is of comparable size, leading to a net gain of order $N$). One can hope, therefore, that conjugating with $T(\eta)$ we can preserve condensation and, at the same time, decrease the energy to make up for the difference between (\ref{eq:fact}) and the true ground state energy~(\ref{eq:gs2}).

\section{Renormalized Excitation Hamiltonian}

We introduce the renormalized excitation Hamiltonian
\begin{equation}\label{eq:cGN} \cG_N = T^* (\eta) \cL_N T(\eta) = T^* (\eta) U_N H_N U_N^* T(\eta) : \cF_+^{\leq N} \to \cF_+^{\leq N} \end{equation}
with $\eta$ defined as in (\ref{eq:etap}).  The next proposition was proven in \cite{BBCS1}. 
\begin{prop}\label{prop:GN1}
Let $V \in L^3 (\bR^3)$ be non-negative, spherically symmetric and compactly supported. Let the coupling constant $\kappa > 0$ be small enough. Let $\cG_N$ be defined as in (\ref{eq:cGN}). Then we can write 
\begin{equation}\label{eq:cGN1} \cG_N = 4 \pi \frak{a}_0 N + \cH_N + \delta_{\cG_N} \end{equation}
where \[ \cH_N = \sum_{p \in \Lambda^*_+} p^2 a_p^* a_p + \frac{\kappa}{2N}\sum_{\substack{p,q \in \Lambda^*_+ , r \in \Lambda^* : \\ r \not = -p , -q}} \widehat{V} (r/N) a_{p+r}^* a_q^* a_{p} a_{q+r} \]
is the restriction of (\ref{eq:HN-bog}) to $\cF_+^{\leq N}$ and where the remainder operator $\delta_{\cG_N}$ is such that, for all $\alpha > 0$ there exists $C > 0$ with 
\begin{equation}\label{eq:prop-GN} \pm \delta_{\cG_N} \leq  \alpha \cH_N + C \kappa (\cN_+ + 1) \end{equation}
as an operator inequality on $\cF_+^{\leq N}$. 
\end{prop}
To prove (\ref{eq:cGN1}) we apply (\ref{eq:TbT}) to the operators $\cL^{(j)}_N$, $j=0,2,3,4$ in (\ref{eq:cLNj}). It is clear that we will generate terms that are not normally ordered (for this simplified discussion, ignore the remainders $d_p, d_p^*$). To restore normal order, we generate terms of lower order in creation and annihilation operators. Specifically, conjugating $\cL^{(2)}_N$ we generate new constant terms while conjugating $\cL^{(4)}_N$ we generate new quadratic and new constant contributions. The choice (\ref{eq:etap}) of $\eta$ guarantees that, on the one hand, the combination of old and new constant terms reproduces, up to an error of order one, the correct ground state energy (\ref{eq:gs2}) and, on the other hand, that there is a cancellation among quadratic terms that allows us to bound everything in terms of $\cH_N$ and $\cN_+$ (as indicated in (\ref{eq:prop-GN})).  

Noticing that, on $\cF_+^{\leq N}$, the kinetic energy operator $\cK$ is gapped, we find $\cN_+ \leq C \cK \leq C \cH_N$. The bound (\ref{eq:prop-GN}) implies therefore that, if $\kappa > 0$ is small enough, 
\begin{equation}\label{eq:ineq} C \cN_+ - C \leq \frac{1}{2} \cH_N - C \leq \cG_N - 4\pi \frak{a}_0 N \leq C (\cH_N + 1) \, . \end{equation}
Hence, if the $N$-particle wave function $\psi_N \in L^2 (\Lambda)^{\otimes_s N}$ is such that 
$\langle \psi_N , H_N \psi_N \rangle \leq 4 \pi \frak{a}_0 N + \zeta$, then we can write $\psi_N = U^*_N T(\eta) \xi_N$, where the excitation vector $\xi_N = T^* (\eta) U_N \psi_N \in \cF_+^{\leq N}$ is such that 
\begin{equation}\label{eq:cond1} \langle \xi_N , \cN_+ \xi_N \rangle \leq  C \langle \xi_N , \cH_N \xi_N \rangle \leq C (\zeta+1) \, . \end{equation}
It is interesting to remark that (\ref{eq:cond1}) implies Bose-Einstein condensation in the sense of (\ref{eq:cond0}), since
\begin{equation}\label{eq:cond2} \begin{split} 1 - \langle \ph_0 , \gamma_N \ph_0 \rangle &= 1 - \frac{1}{N} \langle \psi_N, a^* (\ph_0) a(\ph_0) \psi_N \rangle = \frac{1}{N} \langle U_N \psi_N, \cN_+ U_N \psi_N \rangle  \\ &= \frac{1}{N} \langle \xi_N , T^* (\eta) \cN_+ T(\eta) \xi_N \rangle \leq \frac{C}{N} \langle \xi_N , (\cN_+ + 1) \xi_N \rangle \leq \frac{C(\zeta+1)}{N} \end{split} \end{equation}
where we used the rules (\ref{eq:rules}) and the bounds (\ref{eq:TNKT}) and (\ref{eq:cond1}) (it is then easy to check that (\ref{eq:cond2}) implies $\gamma_N \to |\ph_0 \rangle \langle \ph_0|$ first of all in the Hilbert-Schmidt topology but then also with respect to the trace norm). Eq. (\ref{eq:cond2}) improves (\ref{eq:cond0}) (in the case of small $\kappa$) by giving a precise and optimal bound on the rate of the convergence of the one-particle density matrix.

We can derive stronger bounds on the excitation vector $\xi_N$ associated with a normalized $N$-particle wave function $\psi_N \in L^2 (\Lambda)^{\otimes_s N}$ if, instead of imposing the condition
$\langle \psi_N, H_N \psi_N \rangle \leq 4 \pi \frak{a}_0 N + \zeta$, we require $\psi_N$ to belong to the spectral subspace of $H_N$ associated with energies below $4\pi \frak{a}_0 N + \zeta$. The proof of the next lemma can be found in \cite[Section 4]{BBCS3}.
\begin{lemma}\label{lm:apri}
Let $V \in L^3 (\bR^3)$ be non-negative, spherically symmetric and compactly supported. Let the coupling constant $\kappa > 0$ be small enough. Let $\psi_N \in L^2 (\Lambda)^{\otimes_s N}$ be normalized and such that $\psi_N = {\bf 1}_{(-\infty ; E_N + \zeta]} (H_N) \psi_N$ where ${\bf 1}_{I}$ indicates the characteristic function of the interval $I \subset \bR$. Then $\psi_N = U_N^* T (\eta) \xi_N$, where the excitation vector $\xi_N = T^* (\eta) U_N \psi_N \in \cF_+^{\leq N}$ is such that 
\begin{equation}\label{eq:apri} \left\langle \xi_N, \left[ (\cN_+ + 1)^3 + (\cN_+ + 1) ( \cH_N + 1) \right] \xi_N \right\rangle \leq C (1 + \zeta^3) \end{equation}
uniformly in $N$.   
\end{lemma}
 
With Lemma \ref{lm:apri} we can go back to the renormalized excitation Hamiltonian and 
we can show that several terms contributing to $\cG_N$ are negligible, in the limit of large $N$, 
on low-energy states. The result is the next proposition, whose proof is given in \cite[Section~7]{BBCS3}.
\begin{prop}\label{prop:43}
Let $V \in L^3 (\bR^3)$ be non-negative, spherically symmetric and compactly supported. Let the coupling constant $\kappa > 0$ be small enough. Let $\cG_N$ be defined as in (\ref{eq:cGN}). Then we can write 
\begin{equation}\label{eq:cGN2} \cG_N = C_{\cG_N} + \cQ_{\cG_N} + \cC_N + \cV_N + \cE_{\cG_N} \end{equation}
where, using the notation $\s_p = \sinh (\eta_p)$ and $\g_p = \cosh (\eta_p)$, 
\begin{equation}\label{eq:cVN} 
\begin{split}
\cC_N &= \frac{\kappa}{\sqrt{N}} \sum_{\substack{p,q \in \Lambda^*_+ : \\ q \not = -p}} \widehat{V} (p/N) \left[ b_{p+q}^* b_{-p}^* (\g_q b_q + \s_q b^*_{-q})  + \text{h.c.} \right] \, ,  \\
\cV_N &= \frac{\kappa}{2N} \sum_{\substack{p,q \in \Lambda^*_+, r \in \Lambda^*: \\ r \not = -p, -q}} \widehat{V} (r/N) \, a_{p+r}^* a_q^* a_p a_{q+r} \, , \end{split} \end{equation}
and where 
\[ \begin{split} C_{\cG_N} =& \;\frac{(N-1)}{2} \kappa \widehat{V} (0) +\sum_{p\in\Lambda^*_+}p^2\sigma_p^2+\kappa\widehat{V}(p/N)\left(\s_p\g_p +\s_p^2\right)
\\ &+\frac{\kappa}{2N} \sum_{p,q\in\Lambda_+^*}\widehat{V} ((p-q)/N)\sigma_q\gamma_q\sigma_p\gamma_p\, + \frac1N\sum_{p\in\Lambda^*}\Big[p^2 \eta_p^{2} + \frac\kappa{2N}\big(\widehat{V} (\cdot/N)\ast \eta \big)_p \eta_p\Big]\;\\
        & - \frac 1 N \sum_{q \in \Lambda^*} \k \widehat V(q/N) \eta_q  \sum_{p \in \Lambda_+^*} \s_p^2
\end{split} \]
and
\[  \cQ_{\cG_N} =\sum_{p\in\Lambda^*_+}\Phi_p \, b^*_pb_p+\frac{1}{2}\sum_{p\in\Lambda^*_+}\Gamma_p \, (b^*_pb^*_{-p}+b^*_pb^*_{-p})
\]
with 
\[ \begin{split}
  \Phi_p&=(\s_p^2 + \g_p^2) \, p^2 +\kappa\widehat{V}(p/N)\left(\g_p+\s_p\right)^2+\frac{2\kappa}{N}\g_p\s_p\sum_{q\in\Lambda^*}\widehat{V}((p-q)/N) \eta_q\\
  &\quad-(\g_p^2+\s_p^2)\frac{\kappa}{N}\sum_{q\in\Lambda^*}\widehat{V}(q/N) \eta_q \, ,  \\ 
 \Gamma_p&=2p^2\s_p\g_p+\kappa\widehat{V}(p/N)(\g_p+\s_p)^2+(\g_p^2+\s_p^2)\frac{\kappa}{N}\sum_{q\in\Lambda^*}\widehat{V}((p-q)/N) \eta_q\\
 &\quad-2\g_p\s_p\frac{\kappa}{N}\sum_{q\in\Lambda^*}\widehat{V}(q/N)\eta_q \, .
\end{split}
\]
Moreover, we have 
\begin{equation}\label{eq:EN-bd}  \pm \cE_{\cG_N} \leq \frac{C}{N^{1/4}} \left[ (\cN_+ + 1)^3 + (\cN_+ + 1) ( \cH_N + 1) \right] \, . \end{equation}
\end{prop} 
On the r.h.s. of (\ref{eq:cGN2}) we have a constant and a quadratic term that can be easily diagonalized by means of a generalized Bogoliubov transformation. From (\ref{eq:EN-bd}) it follows that the error term $\cE_{\cG_N}$ is negligible on low-energy states. There are, however, still two terms, the cubic term $\cC_N$ and the quartic term $\cV_N$ in (\ref{eq:cVN}), whose contribution to the spectrum cannot 
be easily determined and that are not negligible. This is the main difference between the Gross-Pitaevskii regime that we are considering here and regimes described by the Hamilton operator (\ref{eq:HNbeta}), with parameter $0 < \beta < 1$, that were  considered in \cite{BBCS2}. For $\beta < 1$, the expectation for example of the quartic interaction can be bounded by
\[ \begin{split} \langle \xi , \cV_N \xi \rangle &\leq \frac{\kappa}{2N} \sum_{p,q,r \in \Lambda^*_+} |\widehat{V} (r/N^\beta)|  \, \| a_{p+r} a_q \xi \| \| a_{q+r} a_p \xi \| \\ &\leq  \frac{C}{N} \sum_{p,q,r \in \Lambda^*_+} \frac{|\widehat{V} (r/N^\beta)|}{(q+r)^2} (p+r)^2  \| a_{p+r} a_q \xi \|^2  \leq C N^{\beta-1} \langle \xi , \cN_+ \cK \xi \rangle \end{split} \]
where we used the estimate 
\[ \sup_{q \in \Lambda^*_+} \sum_{r \in \Lambda^*_+}  \frac{|\widehat{V} (r/N^\beta)|}{(q+r)^2} \leq C N^{\beta} \, .  \]
Hence, for all $\beta < 1$, the quartic and, similarly, also the cubic terms on the r.h.s. of (\ref{eq:cGN2}) are negligible in the limit $N \to \infty$ and can be included in the error term $\cE_{\cG_N}$. This means that, for $\beta < 1$, we can read off the spectrum of $\cG_N$ (and therefore, of the initial Hamiltonian $H_N$), diagonalizing the quadratic operator on the r.h.s. of (\ref{eq:cGN2}). This is not the case for $\beta =1$.

\section{Cubic Conjugation}

It is not surprising that there are still important contributions hidden in the cubic and quartic terms on the r.h.s. of (\ref{eq:cGN2}). Already from \cite{ESY} and, more recently, from \cite{NRSo}, it follows that Bogoliubov states, i.e. in our setting states of the form $U_N^* T(\mu) \Omega$ for some $\mu \in \ell^2 (\Lambda^*_+)$, can only approximate the ground state energy up to an error of order one, even after optimizing the choice of the function $\mu$. To go beyond this resolution, we need to conjugate $\cG_N$ with a more complicated unitary operator. Since Bogoliubov transformations are the exponential of quadratic expressions in creation and annihilation operators, the natural guess is to use the exponential of a antisymmetric cubic phase. In fact, a similar approach was introduced by Yau-Yin \cite{YY} to obtain a precise upper bound for the ground state energy of a dilute Bose gas in the thermodynamic limit, correct up to second order, in agreement with the Lee-Huang-Yang formula. In our setting, we consider the operator
\begin{equation}\label{eq:Seta} S(\eta) = e^{A(\eta)} = \exp \left( \frac1{\sqrt{N}} \sum_{\substack{ r\in P_H, v\in P_L }} \eta_r \big[ b^*_{r+v} b^*_{-r} (\gamma_v b_v + \sigma_v b^*_{-v} ) - \text{h.c.} \big]  \right)  \end{equation}
where, as above $\s_v = \sinh (\eta_p), \g_v= \cosh (\eta_p)$, and where we used the notation $P_H = \{ p \in \Lambda^*_+ : |p| > \sqrt{N} \}$ and $P_L = \{ p \in \Lambda^*_+ : |p| \leq \sqrt{N} \}$. Here, $\eta \in \ell^2 (\Lambda^*_+)$ is the same function defined in (\ref{eq:etap}) entering the definition of the Bogoliubov transformation $T(\eta)$. With the operator (\ref{eq:Seta}), we can define a new, twice renormalized, excitation Hamiltonian \begin{equation}\label{eq:cJN} \cJ_N = S^* (\eta) \cG_N S(\eta) = S^* (\eta) T^* (\eta) U_N H_N U_N^* T(\eta) S(\eta) : \cF_+^{\leq N} \to \cF_+^{\leq N} \, .  \end{equation}
To study the operator $\cJ_N$, we start from the decomposition (\ref{eq:cGN2}) of $\cG_N$ and we analyze how conjugation with $S(\eta)$ acts on the different terms. The first remark is that, when we conjugate with $S(\eta)$, the growth of the number of particles and of the energy remains bounded, independently of $N$. More precisely, we show in \cite[Section 4]{BBCS3} that 
\begin{equation}\label{eq:NHN} \begin{split} S^* (\eta) \, (\cN_+ + 1)^m \, S(\eta) &\leq C (\cN_+ + 1)^m \, , \\ S^* (\eta) \, (\cH_N + 1) (\cN_+ + 1) \, S(\eta) &\leq C (\cH_N + 1) (\cN_+ + 1) \, .\end{split} \end{equation} 
In particular, Eq. (\ref{eq:NHN}) implies that the error term $\cE_{\cG_N}$ on the r.h.s. of (\ref{eq:cGN2}) remains negligible, after conjugation with $S(\eta)$. To conjugate the quadratic 
operator $\cQ_{\cG_N}$ with $S(\eta)$, we observe first that 
\[ \pm \left[ \cQ_{\cG_N} , A(\eta) \right] \leq \frac{C}{\sqrt{N}} (\cN_+ + 1)^2 \, . \]
This bound, combined with the expansion
\[ S^* (\eta) \cQ_N S(\eta) = \cQ_N + \int_0^1 ds \, e^{-s A(\eta)} \left[ \cQ_N, A(\eta) \right] e^{sA(\eta)} 
\] and with the  first estimate in (\ref{eq:NHN}), implies that 
\begin{equation}\label{eq:SQS} S^* (\eta) \cQ_N S(\eta) = \cQ_N + \cE_{1} \end{equation}
where the error operator $\cE_{1}$ is such that $\pm \cE_{1} \leq C N^{-1/2} (\cN_+ + 1)^2$. 
To conjugate the cubic term $\cC_N$ on the r.h.s. of (\ref{eq:cGN2}), we compute 
\begin{equation} \label{eq:cCNA} \left[ \cC_N , A(\eta) \right] =  \Theta + \wt{\cE}_2 \end{equation}
where 
\[ \begin{split} \Theta = \; & \frac 2 N \sum_{r \in P_H, v \in P_L} \k \big( \widehat V(r/N) +\widehat V((r+v)/N)\big) \eta_r\\ &\hspace{2.5cm} \times  \Big[ \s^2_v + ( \g_v^2 + \s^2_v)\, b^*_v b_{v}  +\g_v \s_v\,  \big( b_v b_{-v} + b_v^* b_{-v}^*\big)\Big] \end{split}  \]
and 
\[ \pm \wt{\cE}_{2} \leq C N^{-1/2} \left[ (\cN_+ + 1)^3 +  (\cH_N + 1) (\cN_+ + 1) \right]   \, . \]
The term $\Theta$ on the r.h.s. of (\ref{eq:cCNA}) is not small, but it is such that  
\[ \pm \left[ \Theta , A(\eta) \right] \leq C N^{-1/2} (\cN_+ + 1)^2 \, . \]
Hence, expanding to second order, we conclude that 
\begin{equation}\label{eq:SCS} \begin{split} S^* (\eta) \cC_N S(\eta) = \; &\cC_N + \frac 2 N \sum_{r \in P_H, v \in P_L} \k \big( \widehat V(r/N) +\widehat V((r+v)/N)\big) \eta_r\\ &\hspace{2.5cm} \times  \Big[ \s^2_v + ( \g_v^2 + \s^2_v)\, b^*_v b_{v}  +\g_v \s_v\,  \big( b_v b_{-v} + b_v^* b_{-v}^*\big)\Big] \\ & + \cE_{2} \end{split} \end{equation}
where
\[ \pm \cE_{2} \leq C N^{-1/2} \left[ (\cN_+ + 1)^3 + (\cH_N + 1) (\cN_+ + 1)\right]   \, . \]
To compute the action of $S(\eta)$ on the quartic interaction $\cV_N$, we proceed similarly (but here we have to expand one contribution up to third order). We obtain that
\begin{equation}\label{eq:SVS} \begin{split} S^* &(\eta) \cV_N S(\eta) \\ = \; &\cV_N - \frac 1 {\sqrt{N}} \sum_{\substack{r \in P_H\\ v \in P_L }}  \k \widehat{V}(r/N)\, \Big[ b^*_{r+v} b^*_{-r} \big( \g_v b_v + \s_v b^*_{-v} \big) + \text{h.c.} \Big] \\
 & - \frac 1 N \sum_{\substack{r \in P_H\\ v \in P_L}} \k  \big( \widehat V(r/N) +\widehat V((r+v)/N) \big)  \eta_r \Big[ \s_v^2 + (\g^2_v + \s^2_v)\, b^*_v b_{v} + \g_v \s_v\,  \big( b_v b_{-v} + b_v^* b_{-v}^*\big)  \Big] \\
&  + \cE_{3}\end{split} 
 \end{equation}
where $\pm \cE_3 \leq C N^{-1/4} [ (\cN_+ + 1)^3 +  (\cH_N + 1) (\cN_+ + 1) ]$. The proof of  (\ref{eq:SQS}), (\ref{eq:SCS}) and (\ref{eq:SVS}) can be found in \cite[Sect. 8]{BBCS3}. Combining these results with Prop. \ref{prop:43}, we arrive at the following proposition. 
\begin{prop}
Let $V \in L^3 (\bR^3)$ be non-negative, spherically symmetric and compactly supported. Let the coupling constant $\kappa > 0$ be small enough. Let $\cJ_N$ be defined as in (\ref{eq:cJN}). Then we can write 
\begin{equation}\label{eq:cJN1} \cJ_N = C_{\cJ_N} + Q_{\cJ_N} + \cV_N + \cE_{\cJ_N}
\end{equation}
with  
\[ \begin{split}  \label{eq:tildeCN}
       C_{\cJ_N} :=&\;\frac{(N-1)}{2} \kappa \widehat{V} (0) +\sum_{p\in\Lambda^*_+}\Big[p^2\sigma_p^2+\kappa\widehat{V}(p/N)\sigma_p\g_p +\kappa\big(\widehat{V} (\cdot/N)\ast\widehat{f}_{N,\ell}\big)_p\sigma_p^2\Big]\\
       &+ \frac\kappa{2N}\sum_{p,q\in\Lambda_+^*}\widehat{V} ((p-q)/N)\sigma_q\gamma_q\sigma_p\gamma_p + \frac1N\sum_{p\in\Lambda^*}\Big[p^2 \eta_p^{\;2} + \frac\kappa{2N}\big(\widehat{V} (\cdot/N)\ast\eta\big)_p \eta_p\Big]
           \end{split}
 \]
and the quadratic term 
  \begin{equation} \label{eq:Qtilde}
\begin{split}
       \cQ_{\cJ_N} = \sum_{p\in\Lambda^*_+} F_p \, b^*_p b_p + \frac{1}{2} \sum_{p\in\Lambda^*_+}  G_p \, \big(b^*_pb^*_{-p}+b_p b_{-p} \big) 
\end{split}
\end{equation}
with 
 \begin{equation} \label{eq:defFpGp}
\begin{split}
F_p =&\;p^2(\sigma_p^2+\gamma_p^2)+\kappa\big(\widehat{V} (\cdot/N)\ast\widehat{f}_{N,\ell}\big)_p(\gamma_p+\sigma_p)^2;\\
G_p =&\;2p^2\sigma_p\gamma_p+\kappa\big(\widehat{V} (\cdot/N)\ast\widehat{f}_{N,\ell}\big)_p(\gamma_p+\sigma_p)^2
\end{split}
\end{equation}
where $\widehat{f}_{N,\ell}$ is the Fourier series of $x \to f_{N,\ell} (x) = f_\ell (Nx)$. Moreover, the error operator $\cE_{\cJ_N}$ is such that, on $\cF_+^{\leq N}$,  
  \begin{equation}\label{eq:tildeEN}
         \pm \cE_{\cJ_N}  \leq C N^{-1/4} \Big[(\cH_N+1)(\cN_++1) + (\cN_++1)^3\Big]\,.
         \end{equation}
\end{prop}

\section{Diagonalization and Excitation Spectrum}

Comparing (\ref{eq:cJN1}) with the decomposition (\ref{eq:cGN2}) for $\cG_N$, we notice the absence of the cubic term $\cC_N$, achieved through conjugation with the cubic phase $S(\eta)$. The quartic term $\cV_N$ still appears on the r.h.s. of (\ref{eq:cJN1}) but it is non-negative and therefore we do not worry about it. To read off the excitation spectrum, we conjugate $\cJ_N$ with a last Bogoliubov transformation that diagonalize the quadratic operator $\cQ_{\cJ_N}$. For $p \in \Lambda_+^*$, we define 
$\tau_p \in \bR$ such that (remark that, for $\kappa > 0$ small enough, it is clear that the coefficients $F_p, G_p$ defined in (\ref{eq:defFpGp}) satisfy $|G_p / F_p| < 1$) 
\begin{equation}\label{eq:taup} \tanh (2\tau_p) = - \frac{G_p}{F_p}   \, . \end{equation}
Using the coefficients $\tau_p$, we define \[ \cM_N = T^* (\tau) \cJ_N T(\tau) =  T^* (\tau) S^* (\eta) T^* (\eta) U_N H_N U_N^* T(\eta) S(\eta) T(\tau) : \cF_+^{\leq N} \to \cF_+^{\leq N}.\] 
From (\ref{eq:defFpGp}) and with the definition (\ref{eq:etap}) of $\eta$, we find that $\tau \in H^1 (\Lambda^*_+)$ with norm bounded uniformly in $N$. It follows from (\ref{eq:TNKT}) that conjugation with $T(\tau)$ can only increase number of particles and energy by bounded quantities. This makes it easy to control the action of $T(\tau)$. We find (see \cite[Section 5]{BBCS3} for more details) that 
\begin{equation}\label{eq:cor} \begin{split} \cM_N = &\;4\pi (N-1) \frak{a}_N+ \frac{1}{2} \sum_{p \in \Lambda^*_+} \left[ - p^2  - 8\pi \frak{a}_0 + \sqrt{p^4 + 16 \pi \frak{a}_0  p^2} + \frac{(8\pi \frak{a}_0)^2}{2p^2} \right] \\ &+ \sum_{p \in \Lambda^*_+} \sqrt{|p|^4 + 16 \pi \frak{a}_0 p^2} \; a_p^* a_p + \cV_N + \cE_{\cM_N} \end{split} \end{equation}
where  
\[ \pm \cE_{\cM_N} \leq C N^{-1/4}  [ (\cH_N +1) (\cN_+ +1) + (\cN_+ +1)^3 ]  \, . \]
Theorem \ref{thm:main} now follows making use of the min-max principle to compare the eigenvalues of $\cM_N$ (which coincide with those of the initial Hamiltonian (\ref{eq:ham0})) with those
of the quadratic operator
\[ \begin{split} \cD_N = \; &4\pi (N-1) \frak{a}_N +\frac{1}{2} \sum_{p \in \Lambda^*_+} \left[ - p^2  - 8\pi \frak{a}_0 + \sqrt{p^4 + 16 \pi \frak{a}_0  p^2} \right] \\ &+ \sum_{p \in \Lambda^*_+} \sqrt{|p|^4 + 16 \pi \frak{a}_0 \, p^2} \, a_p^* a_p  \end{split}  \] 
applying the a-priori bound (\ref{eq:apri}) to control the contribution of the error term $\cE_{\cM_N}$. The quartic interaction $\cV_N$ can be neglected in the lower bounds because of its positivity. To prove 
that it can be neglected also in the proof of the necessary upper bounds, it is enough to observe that, on the range of the spectral projection ${\bf 1}_{(-\infty ; \zeta]} (\cD_N)$ (spanned by finitely many eigenvectors of $\cD_N$), we have (see \cite[Section 6]{BBCS3}) 
\[ \cV_N \leq C N^{-1} (\zeta + 1)^{7/2} \, \, . \]

{\it Acknowledgements.} The author gratefully acknowledges support from the NCCR SwissMAP and from the Swiss National Foundation of Science through the SNF Grant ``Dynamical and energetic properties of Bose-Einstein condensates''.

\end{document}